\documentclass[aps,prstab,reprint,groupedaddress]{revtex4-1}

\usepackage{amsmath}
\usepackage{graphicx}
\usepackage[english]{babel}

\begin{document}

\title{The Pulse Intensity-Duration Conjecture: Evidence from Free-Electron Lasers}
\author{T.~Seggebrock}
\email[]{thorben.seggebrock@physik.uni-muenchen.de}
\author{I.~Dornmair}
\author{T.~Tajima}
\affiliation{Ludwig-Maximilians-Universit\"at, 85748 Garching, Germany}
\author{G.~Mourou}
\affiliation{Institut Lumi\`ere Extr\^eme, 91761 Palaiseau, France}
\author{F.~Gr\"uner}
\affiliation{Universit\"at Hamburg, 22607 Hamburg, Germany}
\date{\today}

\begin{abstract}
The recent remark by G.\ Mourou and T.\  Tajima (Science 331, 41 (2011)) on the intensity of the driver laser pulse and the duration of the created pulse that higher driver beam intensities are needed to reach shorter pulses of radiation remains a conjecture without clear theoretical reasoning so far. Here we offer its extension to the case of relativistic electron bunches as the laser's radiating medium (free-electron laser). This also bolsters the understanding of the underlying physical principle of the Conjecture. 
\end{abstract}

\pacs{42.65.Re,41.60.Cr}

\maketitle

\section{Introduction}
In the past few years the interest in high field science and the characterization of ultrafast processes has become intense and their relationship has begun to be studied. G.\ Mourou and T.\  Tajima \cite{Mourou:2011vc,*T.-Tajima:2011fk} recently succeeded in relating ultrafast science to high field science through a specific relationship. From many experiments and realizations of lasers and laser-driven radiation sources a pulse intensity-duration Conjecture emerged when the optimal (i.e.,\ shortest) possible data envelopes were taken. It shows an inverse linear dependence of the pulse duration of radiation on the intensity of the driving laser that produces the pulse, covering 15 orders of magnitude. This Conjecture may be stated as: ``To decrease the achievable pulse duration, we must first increase the intensity of the driving laser.'' This is not the same as the converse trivial statement ``to increase the achievable peak intensity of a pulse for a given energy, we must shorten pulse duration.''\cite{Mourou:2011vc,*T.-Tajima:2011fk}.

This Conjecture is based on the observation of properties of different existing laser facilities and experiments ranging from the first laser developed by T.\ Maiman \cite{Maiman:1960wi} to High Harmonic Generation (HHG) from gas \cite{Antoine:1996vs} and HHG from solids \cite{Dromey:2006gh}. Due to the different underlying physical regimes no analytical model has been provided to explain this Conjecture. The concept, however, may be understood as ``matter exhibits nonlinearities when exposed to strong enough laser radiation; manifestly nonlinearities vary depending on the strength of the ``bending'' field (and thus the intensity). The stronger we ``bend'' the constituent matter, the more rigid the ``bending'' force we need to exert; the more rigid the force is, the higher the restoring frequency (or the shorter the time scale) is.''\cite{Mourou:2011vc,*T.-Tajima:2011fk}. 

However, in their Conjecture no theoretical reasoning is attached beyond this and it remains an empirical conjecture. We wish to see if this Conjecture has a more solid theoretical reason behind it. Furthermore, though the Conjecture was applied only to solid-state lasers (and their applications), it is curious to see if this also applies to other types of lasers.

In this paper we investigate (1) if a Free-Electron Laser (FEL) has this scaling and (2) if we can add a more solid reasoning to the Conjecture using the 1D theory of FELs.

The physics of FELs provides a unique opportunity to test if this Conjecture is in fact applicable to lasers not based on solid state materials and operating in quite different fashions from solid state lasers. The Conjecture was derived from various materials, the laser media themselves, gas targets, solid state targets, etc. On the other hand, FELs provide one specific but completely different material (relativistic electrons in magnetic undulators) that can vary over a wide range of parameters. This allows us to use a single analytical model that governs the driver intensity and the duration of the pulse in order to discuss the Conjecture.

\section{Test of the pulse intensity-duration conjecture}
The Conjecture shows an inverse linear relationship between the intensity of the driver and the resulting pulse duration if one compares systems optimized for short pulse durations. The lasers investigated in ref.\ \cite{Mourou:2011vc,*T.-Tajima:2011fk} were all driven by solid-state lasers. We now wish to consider lasers whose emission mechanism is based on relativistic electron bunches as the radiating medium. In other words we will examine if the stiffness of the relativistic electron bunch contributes to the even shorter pulse duration of this type of laser. 

When studying this relationship in the case of an FEL, one has to be careful to clearly distinguish between the different contributing effects and parameters. In the case of typical lasers, like the Ti:sapphire laser, one has the internal peak intensity in the gain medium on the one hand, and the duration of the resulting laser pulse on the other hand. In the case of High Harmonic Generation one uses the laser intensity impinging on the surface and the duration of the resulting radiation pulse emitted by the medium. Despite these differences, all these cases have one thing in common: one can clearly distinguish between the intensity of the driving pulse and the medium creating the final radiation pulse, which is the dictated condition for checking the Conjecture. In the case of a self amplification of spontaneous emission (SASE) FEL, the situation is not identical to the one above. A SASE FEL is driven by the intensity of the electron beam propagating through the undulator, since it is the electron beam that creates the initial radiation pulse. At the same time this very electron beam acts as the medium amplifying the radiation pulse which co-propagates within it. So, when discussing the Conjecture in the context of an FEL, one has to be careful since by changing the intensity of the electron beam, one might also influence the medium.

The driver intensity in the case of an FEL can be discussed in terms of the electron beam intensity given by
\begin{align}
I_{\text{beam}} &=\frac{P_{\text{beam}}}{2\pi\sigma_{r}^{2}},\\
&= \frac{m_{e}c^{2}\gamma I }{2\pi \sigma_{r}}.
\end{align}
Here we use the electron beam power \(P_{\text{beam}}\), the electron mass \(m_{e}\), the speed of light \(c\), the normalized energy \(\gamma\), the current \(I\), and the beam radius \(\sigma_{r}\). This leads to three dependencies when one wants to discuss influences of the driver intensity: first, the normalized energy; second, the current; and third, the beam size.

Increasing the electron energy has two effects. First, the driving intensity is increased, second, the electron beam gets more rigid and results in a harder-to-bend medium. This should lead to shorter pulses of radiation according to the Conjecture. This is closely related to the cases discussed in the Conjecture, where one had to increase the driver intensity in order to excite higher order nonlinearities, resulting in shorter pulses. Changing the current or the cross section of the beam, on the other hand, does not directly scale the ``spring constant'' of the medium and one can, therefore, not expect a scaling in line with the Conjecture.

On the other hand, a typical measure for the pulse duration is given by the Fourier limit, i.e.\ the inverse bandwidth. Although concepts to further reduce the pulse duration in the case of FELs, e.g.\ by using an energy chirped electron bunch as shown by Saldin et al.\ \cite{Saldin:2006in} exist, we do not discuss such scenarios here, since they are not related to the direct scaling discussed in the Conjecture. The duration, i.e.\ the inverse bandwidth or coherence time, of a single radiation spike is given by \cite{Saldin:1998ul}
\begin{equation}
\tau = \frac{\sqrt{\pi}}{\sigma_{\omega}},
\label{eq:t1}
\end{equation}
using the FEL SASE-bandwidth \(\sigma_{\omega}\). The bandwidth is defined by \cite{Kim:1986tb}
\begin{equation}
\sigma_{\omega}(z)=\sqrt{\frac{3\sqrt{3}\rho}{k_{u}z}}\omega_{l},
\label{eq:swz}
\end{equation}
with \(\rho\) being the Pierce-parameter \cite{Bonifacio:1984tf}, \(k_{u}=2\pi/\lambda_{u}\) the undulator wavenumber of an undulator with period length \(\lambda_{u}\), \(z\) the longitudinal position inside the undulator, and \(\omega_{l}\) the resonant frequency.

To minimize the free parameters, we limit our discussion to an FEL operating at the beginning of saturation and replace the longitudinal position with the saturation length approximately given by \cite{Huang:2007jm}
\begin{equation}
L_{\text{sat}} \approx \frac{\lambda_{u}}{\rho},
\end{equation}
resulting in the relation
\begin{equation}
\sigma_{\omega}(L_{\text{sat}}) = \sqrt{\frac{3\sqrt{3}}{2\pi}} \rho \omega_{l}.
\label{eq:swsat}
\end{equation}
Insertion of the Pierce-parameter and the resonant frequency into eq.\ \eqref{eq:swsat} leads to
\begin{equation}
\sigma_{\omega} = \sqrt{6\sqrt{3}\pi} \left(  \frac{I}{I_{A}} \left(\frac{K \operatorname{JJ} \lambda_{u}}{\sqrt{2}2\pi\sigma_{r}}  \right)^{2} \right)^{1/3} \frac{c\gamma}{\lambda_{u}(1+\frac{K^{2}}{2})},
\end{equation}
with  the Alfv\'en current \(I_{A}=17\) kA, \(K\) the undulator parameter, the Bessel function dependent factor \(\operatorname{JJ} =\operatorname{J}_{0}(\zeta) - \operatorname{J}_{1}(\zeta) \) with \(\zeta=K^{2}/(4+2K^{2})\), and the electron rms beam size \(\sigma_{r}\).

As a start we restrict our discussion to the matched beam size of the undulator. This minimizes the number of free parameters, since it only depends on the focusing properties of the undulator and requires no external focusing system. 
Therefore, it can be seen as the ``natural'' beam size of the setup. By depending only on the setup and not on the electron energy it is especially suited for the discussion of the Conjecture, as it will not change when varying the driver intensity. Other choices of the beam size indeed depend on the electron energy and are discussed in the next section.
%The matched beam size can be seen as the ``natural'' beam size of the setup and is especially suited for the discussion of the Conjecture, since, by depending only on the setup, it will not change when varying the driver, i.e.\ the electron beam, intensity by changing the energy in contrast to other possible choices of the beam size. Other choices of the beam size are discussed in the next section.
The matched beam size of a twofold focusing undulator is given by \cite{Reiche:1999uk}
\begin{equation}
\sigma_{r}=\sqrt{\frac{\sqrt{2}\epsilon_{n}}{K k_{u}}},
\label{eq:srmb}
\end{equation}
with the normalized rms emittance \(\epsilon_{n}\). Using this and the expression for the bandwidth above, this results in a pulse duration of
\begin{equation}
\tau = \frac{1}{\sqrt{6\sqrt{3}}} \left( \frac{I_{A}}{I}  \left(  \frac{2\sqrt{\pi\epsilon_{n}\sqrt{2}}}{K^{3/2}\operatorname{JJ}\sqrt{\lambda_{u}}}   \right)^{2} \right)^{1/3} \frac{\lambda_{u}(1+\frac{K^{2}}{2})}{c\gamma}.
\label{eq:t}
\end{equation}
Here the general relationship that the pulse duration drops with increasing electron energy and current, which are directly proportional to the electron beam intensity, can already be seen. We find here a first hint that an FEL shows a scaling supporting the Conjecture. 

The Conjecture expects that only when optimized setups are achieved then the shortest duration of radiation emerges. Thus, we keep the setup parameters \(K\) and \(\lambda_{u}\) as well as the normalized emittance \(\epsilon_{n}\) fixed and assume them to be optimized for a short pulse length.
This leads us to the only free parameters: the current \(I\) and the normalized electron energy \(\gamma\).  

Using these arguments, the coherence time of the created photon pulse is related to the electron beam intensity as
\begin{equation}
\tau \propto \gamma^{-1} \propto I_{\text{beam}}^{-1},
\label{eq:tg}
\end{equation}
when only varying the energy for a fixed current. On the other hand, another dependence emerges
\begin{equation}
\tau \propto I^{-1/3} \propto I_{\text{beam}}^{-1/3},
\end{equation}
for a fixed normalized energy and variation of the current.

\begin{figure}

\includegraphics{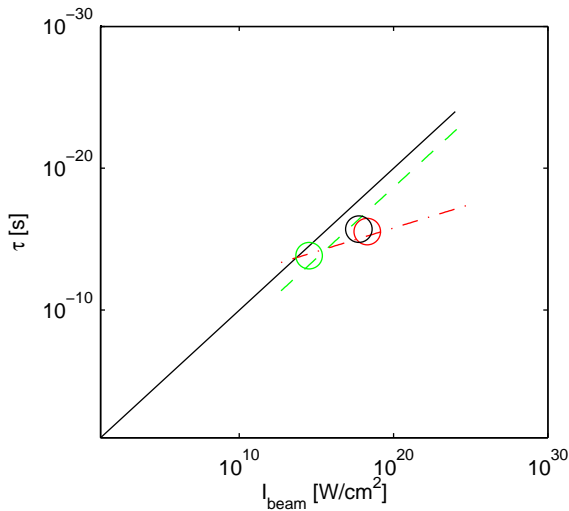}

\caption{(a) Comparison of the Conjecture found by G.\ Mourou and T.\  Tajima (solid black line), the here derived dependence of the coherence time on the beam intensity for a fixed current (dashed green line) and a fixed energy (dash-dotted red line), and the data points of LCLS (red circle), FLASH (green circle) and the X-FEL (black circle).}
\label{fig:conjecture}
\end{figure}

Fig.\ \ref{fig:conjecture} shows the comparison of the Conjecture found by G.\ Mourou and T.\  Tajima, and the derived dependencies of the coherence time on the beam intensity. In the case of the fixed current the dependence is inverse linear as in the Conjecture. This may be understood as follows: assumption of a fixed current represents the situation of a medium that gets harder to bend due to the increasing energy, and results in shorter pulses. Now we see, the Conjecture emerges when we have related the pulse duration and its functional dependence to the energy of the electron beam (i.e.,\ the \(\gamma\) factor), as the latter implies the stiffness of the medium.
The slope of the red line in Fig.\ \ref{fig:conjecture}, i.e.\ the case of a fixed energy and variation of the current, can be related to a simultaneous change of the density of the medium and the driver intensity due to the increasing current (and particle number in the beam) resulting in a different scaling. This is reasonable, as the ``spring constant'' in this case is not directly scaled. This leads to the conclusion that the scaling of the Conjecture is indeed due to the rigidity of the pulse-generating matter, and that high driver intensities are needed to either; reach these rigidities, as in the case of an FEL, or to excite more rigid non-linearities, as in the cases discussed in the original Conjecture.

Besides the different scaling, both cases support the Conjecture, since they lead to the same result: to reduce the pulse duration, one has to increase the driver intensity.

\section{Role of optimization}
An interesting result of the Conjecture is the product of pulse duration and intensity, the coefficient of the graph. It is the constant of the Conjecture, with a value of \(1\) J\,cm\(^{-2}\). We can study this in the case of a fixed current, resulting in a product of coherence time and beam intensity of
\begin{equation}
\tau I_{\text{beam}} = \frac{\pi^{1/3}}{2^{1/6}\sqrt{3\sqrt{3}}} m_{e}c I_{A}^{1/3}\frac{I^{2/3}(1+\frac{K^{2}}{2})}{\epsilon_{n}^{2/3}\operatorname{JJ}^{2/3}\lambda_{u}^{1/3}}.
\end{equation}
Using the parameters \(I=1\) kA, \(\lambda_{u}=1\) cm, \(K=1\), and \(\epsilon_{n}=1\) mm\,mrad, which are reasonable for current FELs, the product equals approximately \(22.5\) J\,cm\(^{-2}\). Lower values of the product are regarded more favorable (or more efficient), since in that case lower beam intensities are needed to reach shorter pulses. In the case of an FEL the product may easily range from 1 J\,cm\(^{-2}\) to 100 J\,cm\(^{-2}\) by adjusting the setup parameters. The fractional power dependence of the product on the current is mainly due to the beam intensity, the scaling with the other parameters like \(\lambda_{u}\), \(K\), and \(\epsilon_{n}\) are more complex and are a result of the competing effects between the coherence time and the beam intensity.

\begin{figure}
\begin{tabular}{c}
\includegraphics{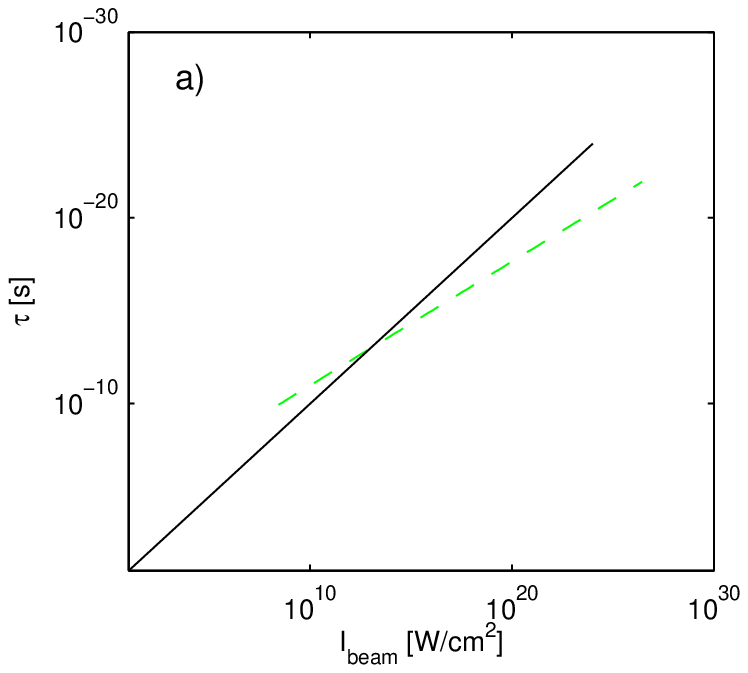}\\
\includegraphics{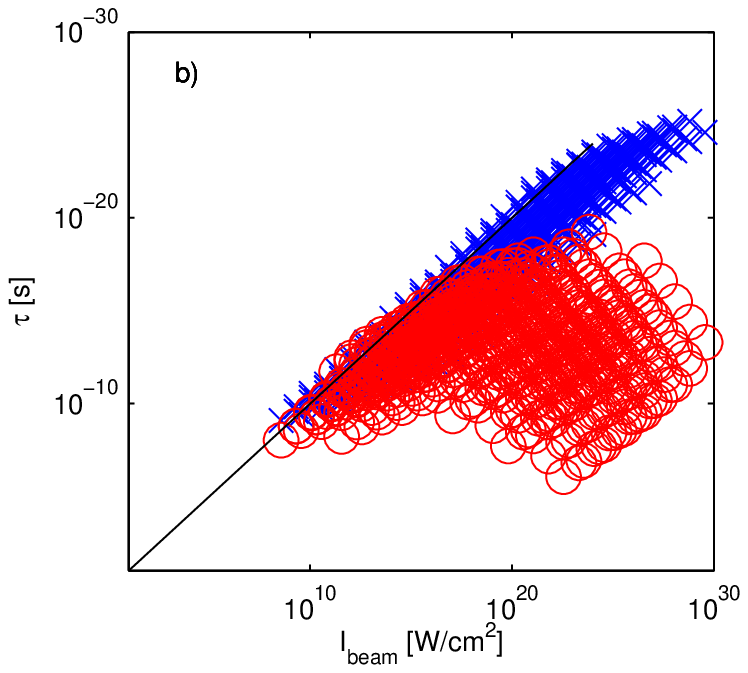}
\end{tabular}
\caption{(a) Dependence of the coherence time on the beam intensity for a nonmatched beam with constant beta function (dashed green line) in comparison to the Conjecture (solid black line).\\
(b) Coherence times for different beam intensities according to Eq.\ \ref{eq:t} (blue crosses), and according to Eqs.\ \ref{eq:t1} and \ref{eq:swsat}  calculated using the fitting formula of M.\ Xie \cite{Xie:1996wa} taking degrading effects into account (red circles). The following parameters were varied: \(\gamma=10^{1}\to10^{12}\), \(\lambda_{u}=10^{-3}\to10^{0}\) m, \(K=10^{-2}\to10^{1}\) and \(I=10^{2}\to10^{6}\) A. The emittance was set to \(\epsilon_{n}=10^{-6}\) m\,rad. The black line is again the Conjecture as a reference.}
\label{fig:conjecture_notmatched}
\end{figure}

Note that often in real FEL facilities the beam size might not be optimized for a short coherence time, but for maximum gain. Therefore, external focusing optics are used to focus the beam to values below the matched beam size. These cases should not line up to the line suggested by the Conjecture, since it governs only the shortest possible pulse duration for a given driver intensity. For the sake of completeness we discuss some other scalings, although they do not cover all the optimizations in current FEL facilities. The beam spot size in  the case of an external focusing system is given by \(\sigma_{r}=\sqrt{\beta\epsilon_{n}/\gamma}\), with \(\beta\) being the beta function of the focusing system, resulting in a different dependence of the coherence time on the beam intensity for a fixed \(\beta\)
\begin{equation}
\tau \propto \gamma^{-4/3} \propto I_{\text{beam}}^{-2/3}.
\end{equation}
Here we again assumed a constant current. In an FEL optimized for short gain length the ideal \(\beta_{\text{opt}}\propto\gamma^{2}\) \cite{Saldin:2004iq} would lead to a more complex dependence. However, this case is not suited for our discussion. This is because on the one hand, this optimization is only valid for a very small parameter range, limited by the ratio between electron beam emittance and radiation wavelength; and on the other hand, the optimization for a short gain length is not part of the scaling discussed in the Conjecture.
The resulting scaling for a fixed \(\beta\) is shown in Fig.\ \ref{fig:conjecture_notmatched}(a). This shows that using an arbitrary beta function in the focusing system is not ideal when aiming for shorter coherence times. The matching of the beam size (Eq.\ \eqref{eq:srmb}) optimizes the pulse length, as can be seen in Eqs.\ \eqref{eq:t} and \eqref{eq:tg}. This is again another manifestation of the Conjecture property. That is, the Conjecture observes that only after optimizations in the laser experiments for the shortest possible pulses, the inverse linear dependence emerges. This feature is also seen in Fig.\ \ref{fig:conjecture_notmatched}(b), where a wide parameter range has been scanned using the derived formulas. The blue domain shows the results according to the 1D theory while the red domain takes degrading effects into account by using the fitting formula of M.\ Xie \cite{Xie:1996wa}.
The parameter scan shows that the upper envelope of the blue domain is represented by the Conjecture line and only degrading effects cause a deviation.
However, if we optimize for the efficiency of lasing instead of the pulse shortness, the scaling line nearly goes to the envelope of the bottom of the blue region in Fig.\ \ref{fig:conjecture_notmatched}(b).

\section{Limitations}
Real FEL facilities are limited in increasing the electron energy due to two effects degrading the FEL performance, quantum diffusion and an additional quasi energy spread due to the emittance of the electron beam. For higher electron energies and therefore higher photon energies the number of emitted photons drops and results in a random emission process not following the classical description called quantum diffusion. The resulting increase of energy spread is given by \cite{Saldin:1996ta}
\begin{equation}
\frac{\mathrm{d} \langle (\Delta\gamma)^{2} \rangle }{\mathrm{d} z} = \frac{7}{15} r_{e} \lambdabar_{c} \gamma^{4} K^{2} k_{u}^{3} \operatorname{F}(K),
\end{equation}
with the classical electron radius \(r_{e}\), the reduced compton wavelength \(\lambdabar_{c}\) and the function
\begin{equation}
\operatorname{F}(K) = 1.2 K + \frac{1}{1+1.33K+0.40K^{2}}.
\end{equation}
This increase of energy spread leads to a violation of the requirement \(\sigma_{\gamma}/\gamma < \rho\) when increasing the electron energy. However, this can be counteracted by reducing the gain length, e.g.\ by using higher currents, and consequently pushing the point of violation towards higher beam intensities, widening the region of applicability of the ideal 1D theory.

The beam emittance can be treated as the source of an additional longitudinal energy spread due to the transverse motion of the electrons. Requiring \(\sigma_{\gamma}/\gamma < \rho\) results in the emittance limit \cite{Reiche:1999uk}
\begin{equation}
\epsilon_{n} \ll \frac{4\rho\beta\gamma\lambda_{r}}{\lambda_{u}}.
\end{equation}
For the matched beam size this limit drops as \(1/\gamma\). Even state-of-the-art accelerators can already reach a normalized emittance on the order of \(10^{-8}\) m\,rad, allowing for the use of the 1D model over several orders of magnitude in beam intensity.

\section{Conclusion}
Using the 1D FEL theory, we have found the inverse linear relation emerges between the coherence time of an FEL pulse and the electron beam intensity upon the optimization for the shortest possible pulse duration. This in fact agrees with the pulse intensity-duration Conjecture found by G.\ Mourou and T.\  Tajima. In the case of a fixed current and a variable energy the dependence is the same inverse linear dependence as in the Conjecture. This result allows for the expansion of the Conjecture from the original form applied to solid state lasers to the completely different regime of Free-Electron Lasers. While the lasers discussed in the Conjecture are based on nonrelativistic drivers interacting with nonrelativistic as well as relativistic materials, the FEL is highly relativistic with still the same \(\tau\)-\(I_{\text{beam}}\) dependence. This implies that the underlying physics behind the Conjecture is neither dependent on specific material properties, nor the distinction between nonrelativistic or relativistic dynamics; but rather it is derived from the rigidity of the radiating material in general and the required intensity of its driver. 

In a more general view, FELs can be regarded as generators of radiation with arbitrary wavelengths, especially wavelengths much shorter than the generating medium itself, which is possible only via self-modulation of this medium. In order to reach shorter and shorter wavelengths (and thus shorter pulse durations), the relativistic gamma factor must be increased. The corresponding increase of the electron mass reduces the coupling between the medium and the self-amplified radiation. This in turn is compensated by an increase of the interaction time between the medium and the radiation field needed to reach saturation, hence the amount of energy transferred from the medium to the field is independent of gamma. This is the reason why the FEL-gain bandwidth increases with gamma, and thus, the FEL-relation between electron beam intensity and resulting pulse duration follows the Conjecture. Only degrading effects as mentioned above cause a deviation from that.

The conclusion we find here on the \(\tau\)-\(I_{\text{beam}}\) dependence is the same as in the case of the Conjecture as originally claimed in 
\cite{Mourou:2011vc,*T.-Tajima:2011fk} - to create shorter pulses, more intense beams are needed. This paper can, therefore, be taken as an extension and proof of the Conjecture found by G.\ Mourou and T.\  Tajima for relativistic electron bunches as the radiation source. An effort to reach for the Schwinger intensity \cite{Bulanov:2003el,Kando:2009be} already hints that this Conjecture extends far into the higher intensities. At the same time, spurred by this Conjecture more novel approaches for the detection of even less than attosecond pulses are beginning \cite{Ipp:2011du}. These are examples of the useful and guiding role this Conjecture is playing for the future. 

The underlying mechanism is identified as the rigidity of the radiating medium. This finding leads us to further contemplate a potential extension of the Conjecture. Can it be the case that apart from the shortening of photon pulses, in order to shorten the bunch of charged particles, we have to increase the intensity of the bunch? Clearly, the inverse is already true, but the above is not trivial and never mentioned. It is encouraging from our present investigation to see if the Conjecture extends beyond the realm of lasers in general to even charged particle beams. Though we leave detailed investigation of this question for the future, we see at least a trend toward what the extended Conjecture to charged particle beams could imply.

\begin{acknowledgments}
The authors enjoyed the stimulating and supportive discussions with F.\ Krausz, E.\ Goulielmakis, M.V.\ Yurkov, and E.A.\ Schneidmiller. The work is supported by DFG Cluster of Excellence MAP (Munich Centre for Advanced Photonics).
\end{acknowledgments}

\bibliography{ConjectureFEL.bib}

\end{document}